\documentclass[twocolumn,aps,groupedaddress,nofootinbib]{revtex4}
\usepackage{graphicx}
\usepackage{dcolumn}
\usepackage{bm}
\usepackage{amsmath,amssymb,amsfonts}
\usepackage{color}
\usepackage{braket}
\usepackage{mathtools}

\usepackage{amsmath} 

\begin{document}

\title{Terahertz detection based on nonlinear Hall effect without magnetic field}

\author{Yang Zhang}
\author{Liang Fu}
\affiliation{Department of Physics, Massachusetts Institute of Technology, Cambridge, Massachusetts 02139, USA}

\begin{abstract}
We propose a method for broadband long-wavelength photodetection using the nonlinear Hall effect in non-centrosymmetric quantum materials. The inherently quadratic relation between transverse current and input voltage at zero magnetic field is used to rectify the incident terahertz or infrared electric field into a direct current, without invoking any diode. Our photodetector operates at zero external bias with fast response speed and has zero threshold voltage. 
Remarkably,  the intrinsic current responsivity due to Berry curvature mechanism is a material property independent of the incident frequency or the scattering rate, which can be evaluated from first-principles electronic structure calculations. We identify the Weyl semimetal NbP for terahertz photodection with large current responsivity reaching $\sim 1$A/W without external bias.
\end{abstract}

\maketitle

Quantum materials is one of the most important fields in condensed matter physics. While at a fundamental level the properties of all materials are governed by quantum mechanics, quantum materials are distinctive in that they exhibit novel electrical and optical properties originating from the quantum nature of electron wavefunction. The complexity and richness of quantum wavefunction underlie new phases of matter and emergent properties, as revealed through the lens of geometry \cite{ahereview}, topology \cite{kanemele} and entanglement \cite{lucilebalents}. The rise of quantum materials not only advances our fundamental understanding of solids but also opens exciting opportunities for inventing new technologies. 


Terahertz technology is a fast-growing field with wide-ranging applications. Since terahertz waves can transmit through many materials, terahertz spectroscopy and imaging can be used to identify concealed objects for security check, detect chemical composition and material defects for quality control, and examine biological tissues in medical imaging. Terahertz sensors can help autonomous vehicles to identify remote objects  under foggy conditions. Terahertz-band communication is regarded as a key enabling technology for next-generation wireless network. Despite the enormous potential, fast, sensitive and broadband terahertz detection at room temperature has been a technology challenge.  Conventional THz detectors, such as Golay cells, bolometers and pyroelectric sensors, rely on thermal response and therefore usually require cryogenic cooling and have slow operation speed. Alternatively, Schottky diodes in conjunction with antennas are used to convert high-frequency electric fields signals to DC. However, they typically have cutoff frequencies below 1THz. 

In this work we propose a method of terahertz detection based on the intrinsic nonlinear Hall effect in quantum materials. This effect was predicted \cite{Sodemann} and observed \cite{Ma,Mak} in inversion-breaking crystals without the need of magnetic field, where an applied electric field induces a {\it transverse} current to the second order (see also Refs.\cite{Moore, Spivak}). Our idea here is to use this {\it quadratic} transverse current-voltage characteristic of quantum materials to rectify and detect terahertz radiation directly, without invoking any junction region. This ``Hall rectifier'' \cite{Isobe} has a number of important advantages. It is an intrinsic full-wave rectifier with a large attainable responsivity at zero bias and a high cutoff frequency which range from tens to hundreds of THz. It features a very short response time down to $ps$. Due to its simple device architecture, Hall rectifiers can be easily fabricated, integrated on-chip and potentially mass-produced.

Remarkably, we show that the current responsivity due to the intrinsic second-order Hall effect is independent of the frequency and intensity of incident radiation, and it is an intrinsic material property determined by electronic band structure and Berry curvature dipole \cite{Sodemann}. This finding enables us to predict the right materials for high-sensitivity terahertz detection using first-principles calculations. We identify a number of topological semimetals with anisotropic energy dispersion, in which the intrinsic current responsivity can reach  $1$A$/$W without external bias and over a very broad range of frequency.

Unlike photogalvanic effects at optical frequencies due to interband transitions \cite{sipe2000second,de2017quantized,morimoto2016topological,chan2016chiral,ishizuka2016emergent,yang2017divergent,young2012first,parker2019diagrammatic,dipole,osterhoudt2019colossal,rees2020helicity,ni2020giant,ma2019nonlinear,ma2017direct,konig2017photogalvanic}, the photoresponse of metals and degenerate semiconductors at (sub-)terahertz frequencies is largely governed by intraband proccess involving electrons near the Fermi surface only.
Importantly, since the dispersion relation of Bloch electron in nonmagnetic materials necessarily  satisfies $\epsilon_{\bm{k}}=\epsilon_{-\bm{k}}$ due to time reversal symmetry, the second-order response of noncentrosymmetric materials is an inherently quantum-mechanical effect arising from the inversion asymmetry of electron wavefunction within the unit cell. Therefore, quantum materials having complex Bloch wavefunctions, such as topological surface states and inversion-breaking topological metals, generally exhibit large second-order response. 

The nonlinear Hall effect is a type of second-order response which can be viewed as current-induced anomalous Hall effect. The anomalous current due to Berry curvature ${\bf \Omega}_{\bf k}$ is given by
\begin{eqnarray}
{\bf j}^A= \frac{e^2}{\hbar} \int \frac{d^d \bf k}{(2\pi)^d}  f({\bf k}) \left( {\bf E} \times {\bf \Omega}_{\bf k}  \right)
\end{eqnarray}
where $E$ is the applied electric field and $f({\bf k})$ is the distribution function in the presence of the field.
To first order in $E$, the anomalous current is obtained by replacing $f$ with the equilibrium Fermi-Dirac distribution $f^0({\bf k})= 1/(e^{\beta (\epsilon_{\bf k}-\mu)}+1)$.
In time-reversal-invariant and inversion-breaking systems, Berry curvature can be nonzero, but the condition  ${\bf \Omega}_{\bf k}=-{\bf \Omega}_{-\bf k}$  leads to zero linear-response anomalous current due to the cancellation of contributions from Bloch electrons at $\pm \bf k$.

On the other hand, the anomalous current can be nonzero when we extend the analysis to the second order in $E$. In the current-carrying state, the distribution function $f$ differs from equilibrium, and to first-order in $E$, is given by
\begin{eqnarray}
f({\bf k}) = f^0({\bf k}) +  \frac{e {\bf E} \tau}{\hbar} \cdot (\partial_{\bf k} f^0),
\end{eqnarray}
where $\tau$ is the scattering rate in Drude transport.
Then, the anomalous current at second order is obtained
\begin{eqnarray}
j^A_a =\left( \epsilon_{ab\lambda}  \frac{e^3  \tau B_{\lambda c}}{\hbar^2}  \right) E_b E_c  \equiv \chi_{abc} E_b E_c
\end{eqnarray}
where the nonlinear Hall conductivity $\chi_{abc}$  is a third-rank tensor that is antisymmetric in indices $a,b$. $B_{\lambda c}$ is Berry curvature dipole, an intrinsic property of Bloch electron wavefunctions defined by \cite{Sodemann}
\begin{eqnarray}
B_{\lambda c} =\int \frac{d^d \bf k}{(2\pi)^d}
    f^0 ({\bf k}) \frac{\partial \Omega_{\lambda}}{\partial k_c  }.
\end{eqnarray}
$B$ is allowed by symmetry in noncentrosymmetric crystals and surfaces having certain point groups (see below).
For example, in materials with a polar axis, large $B$ has been found due to the presence of titled Dirac or Weyl points in the band structure \cite{Sodemann,Zhang,Dzsaber,zhang2018electrically,Brink,You,TMD2,zhou2020highly,pantaleon2020tunable}.
When $B$ is nonzero and in the presence of an applied electric field, the imbalance in the probability of occupation $f({\bf k}) \neq f(-{\bf k})$ leads to  a net Berry curvature, which then induces a second-order anomalous current in the transverse direction.

\begin{figure}
\includegraphics[width=\hsize]{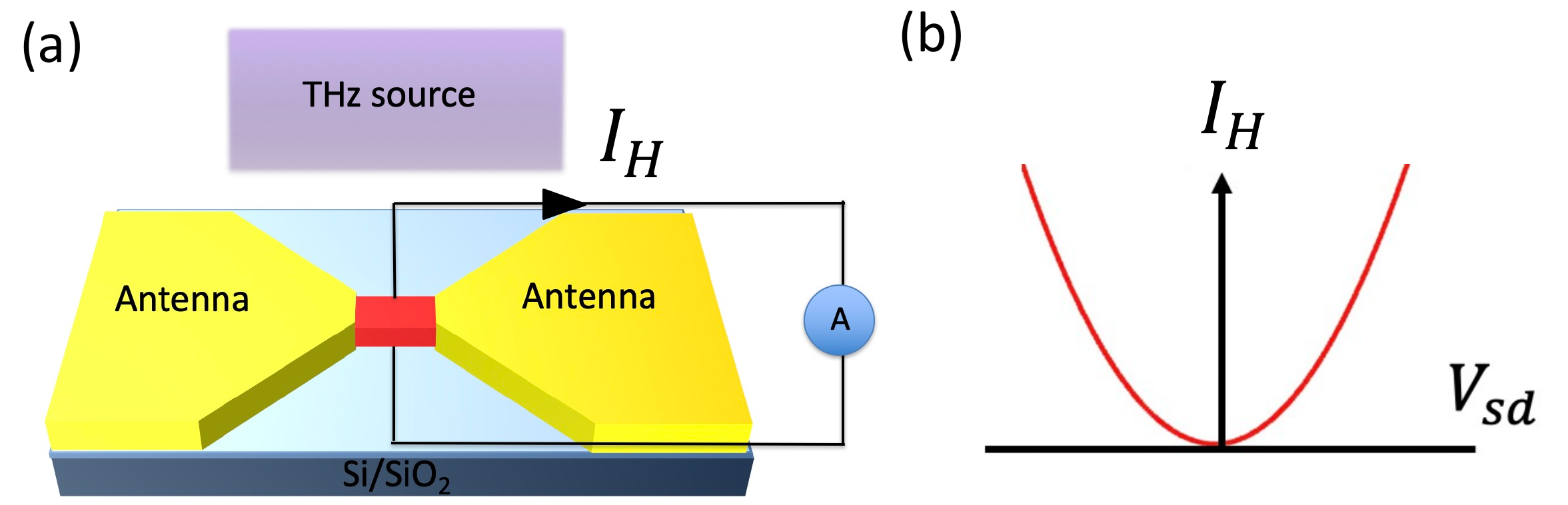}
\caption{
(a) Long wavelength photodetection based on nonlinear Hall effect in quantum materials. The oscillating terahertz field collected by bowtie antenna induces a direct current in the transverse direction via the intrinsic second-order response of the quantum material itself.
(b) Current-Voltage characteristic of nonlinear Hall rectifier. The quadratic relation enables full-wave rectification of AC input at zero external bias, with zero threshold voltage and zero dark current.
}
\label{fig1}
\end{figure}

The inherently quadratic relation between the transverse  current and input voltage due to the nonlinear Hall effect is ideal for rectifying AC input ${\mathbf E}(t)={\mathbf E}\cos(\omega t)$, including both the positive and negative half cycles. At low frequency $\omega \ll 1/\tau$, the time-dependent transverse current is simply $j^A|_{E(t)} \propto \cos^2(\omega t)$, which is a sum of DC and $2\omega$ components with equal weight.  Thus, the DC output $j^0$, or the time average of $j^A(t)$ over a full cycle, is given by $j^0=\frac{1}{2}j^A=\frac{1}{2} \chi E E$.
Unlike diodes, our Hall rectifier does not involve any junction or barrier structure, but relies on the inherent nonlinearity of a homogeneous quantum material. Since the carriers do not need to overcome a barrier to conduct a current, Hall rectifier has a zero threshold voltage and operates without the need of a finite bias, which is ideally suited for detecting small signals in terahertz frequency. Since both positive and negative voltages generate transverse current in the same direction, dark current is eliminated.
Furthermore, as we show below, large nonlinear Hall effect can be found in metals or semimetals, in which the high carrier mobility enables fast device speed.

Generally, at frequencies well below the threshold for interband transition, the second-order direct current due to Berry curvature dipole mechanism shows a Drude-like response \cite{Moore, Spivak, Sodemann}:
\begin{eqnarray}
j^0_a =  
(\frac{1}{1+ \omega^2 \tau^2}) \frac{\chi_{abc}}{2} E_b E_c.
\end{eqnarray}
where the frequency dependence comes from the change of distribution function to the first order of the oscillating electric field. Note that for time-reversal-invariant systems, the nonlinear Hall effect and more generally, linear photogalvanic effect, vanish in the absence of dissipation $\tau^{-1} \rightarrow 0$, because time reversal symmetry forbids a direct current \cite{morimoto2018nonreciprocal}. At frequencies $\omega \gg \tau^{-1}$, the photocurrent $j^0$ from intraband process decreases as $1/\omega^2$.  However, in reality, even in the case of good metals, the scattering rate at room temperature is at least a few tens of THz. Therefore,  the nonlinear Hall effect of a homogeneous quantum material can be used as a replacement of diodes to rectify incident waves in a very broad range of frequency up to at least 10 THz, and the output direct current is linearly proportional to the incident power.

An important parameter for terahertz detectors is current responsivity, which measures the electrical output per optical input.
In our device shown in Fig.1, the antennas collect the incident radiation and create strongly enhanced THz field in the active region of Hall rectifier, which generates the photoresponse current in the transverse direction. The power absorbed by the Hall rectifier $P=\sigma_{ab}(\omega) E_a E_b S$, where $S=LW$ is the area of the Hall rectifier with length $L$ and width $W$. $\sigma$ is the linear-response AC conductivity given by
\begin{equation}
\sigma_{a b}(\omega) =\frac{e^2 D_{ab}}{\hbar^2} \frac{ \tau}{(1+\omega^2 \tau^2)}
\end{equation}
within Drude transport theory, where the Drude weight $D$ is defined by
\begin{equation}
D_{ab}= \int \frac{d^d k}{(2\pi)^d}
\frac{\partial E}{\partial k_{a}} \frac{\partial E}{\partial k_{b}}\left(-\frac{\partial f_{0}}{\partial E}\right).
\end{equation}
The output current per absorbed power defines the current responsivity of our Hall rectifier:
\begin{eqnarray}
R=\frac{I_H}{P}=\frac{1}{W} \frac{j^0 }{\sigma_{ab} E_a E_b }.
\end{eqnarray}
Note that $R$ is inversely proportional to the width of Hall rectifier, so it is advantageous to use a narrow channel.

Due to the quadratic dependence of transverse current $j^0$ on THz field, $R$ is inherently independent of incident power, i.e., our terahertz detector exhibits a linear response over a large range of optical power. Moreover, since both the nonlinear Hall conductivity and linear response conductivity exhibit Drude behavior with the same $\tau/(1+\omega^2 \tau^2)$ dependence, $R$ is independent of frequency and scattering rate, regardless of the value of $\omega\tau$. Therefore the current responsivity due to the intrinsic nonlinear Hall effect is entirely determined by the Berry curvature dipole $B$ and the Drude weight $D$. $B$ is a property of Bloch electron wavefunction, and $D$ a property of energy dispersion.  This is an important result of our work. As both $B$ and $D$ are material properties attainable from first-principles electronic structure calculations, the intrinsic current responsivity $R$ of quantum materials can be predicted without any tuning parameter. This enables computational materials design of long-wavelength photodetectors.

\begin{figure}
\includegraphics[width=\hsize]{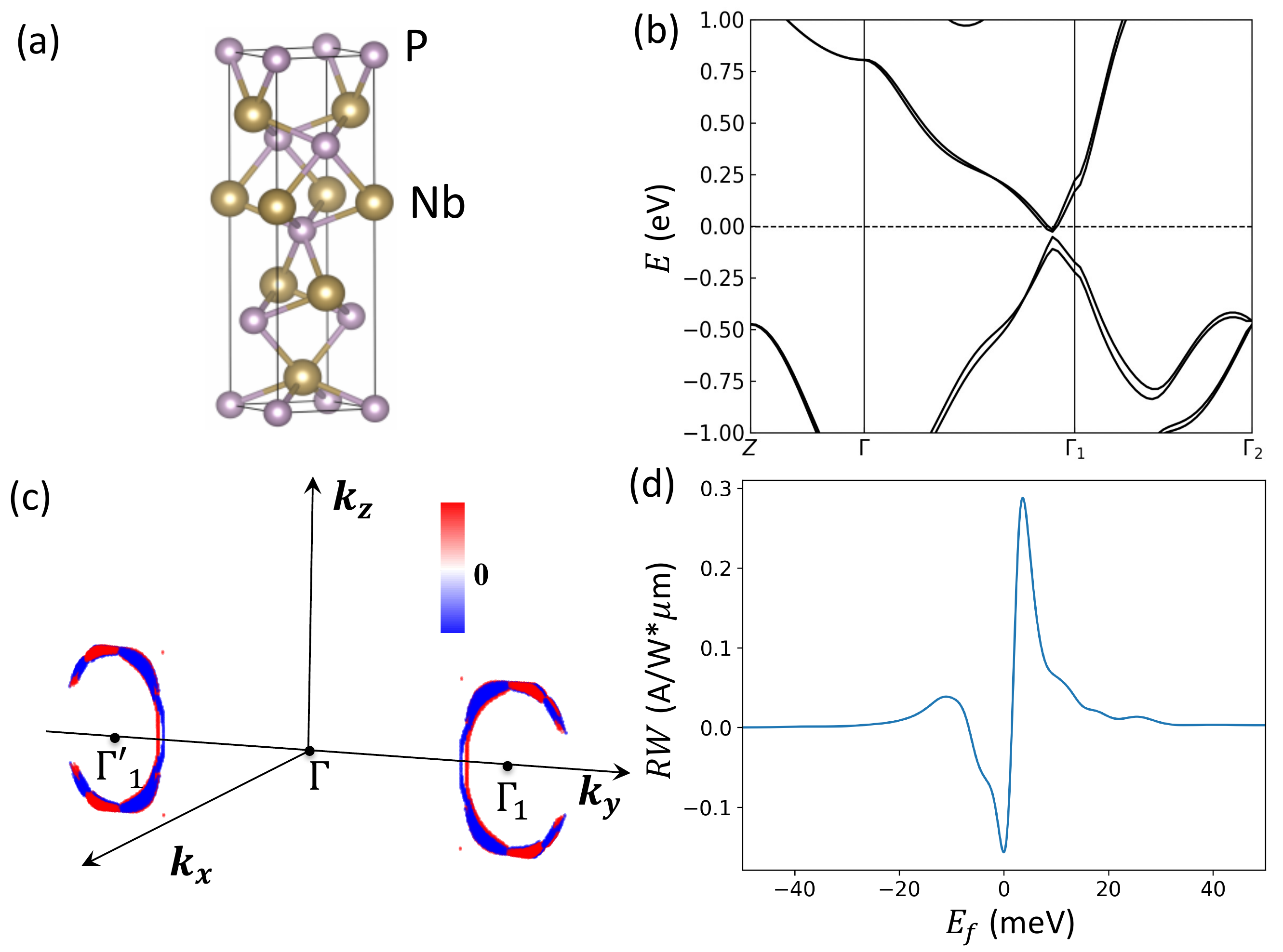}
\caption{
(a) Lattice structure of NbP;
(b) Band structure of NbP, which shows a narrow band inversion around $\Gamma_1$;
(c) 3D momentum resolved Berry curvature dipole;
(d) Current responsivity as a function of chemical potential.
}
\label{fig2}
\end{figure}

To search for suitable quantum materials suitable for Hall rectifier, we note that large Berry curvature dipole appears in the vicinity of band crossings with anisotropic energy dispersion. Therefore inversion-breaking Weyl semimetals such as TaAs, TaP, NbP and NbAs \cite{bansil2016colloquium,hasan2017discovery,weng2015weyl,armitage2018weyl} are good candidates. Our first-principles calculations reveal that among these four materials, NbP has the largest current responsivity. The NbP family of Weyl semimetals has point group $C_{4v}$ with four fold rotation axis in the z direction and two vertical mirror planes $\mathcal{M}_x$, $\mathcal{M}_y$. Thus the only nonzero elements of the Berry curvature dipole tensor are $B_{yx} = -B_{xy}$. Hence, the photoresponse is largest when the incident electric field lies within $xy$ plane and the photocurrent along the $z$ direction is measured. In this configuration, the intrinsic current responsivity is given by
\begin{eqnarray}
R = \frac{1}{W} \frac{e B_{xy}}{2 D_{xx}}.
\end{eqnarray}

\begin{figure}
\includegraphics[width=\hsize]{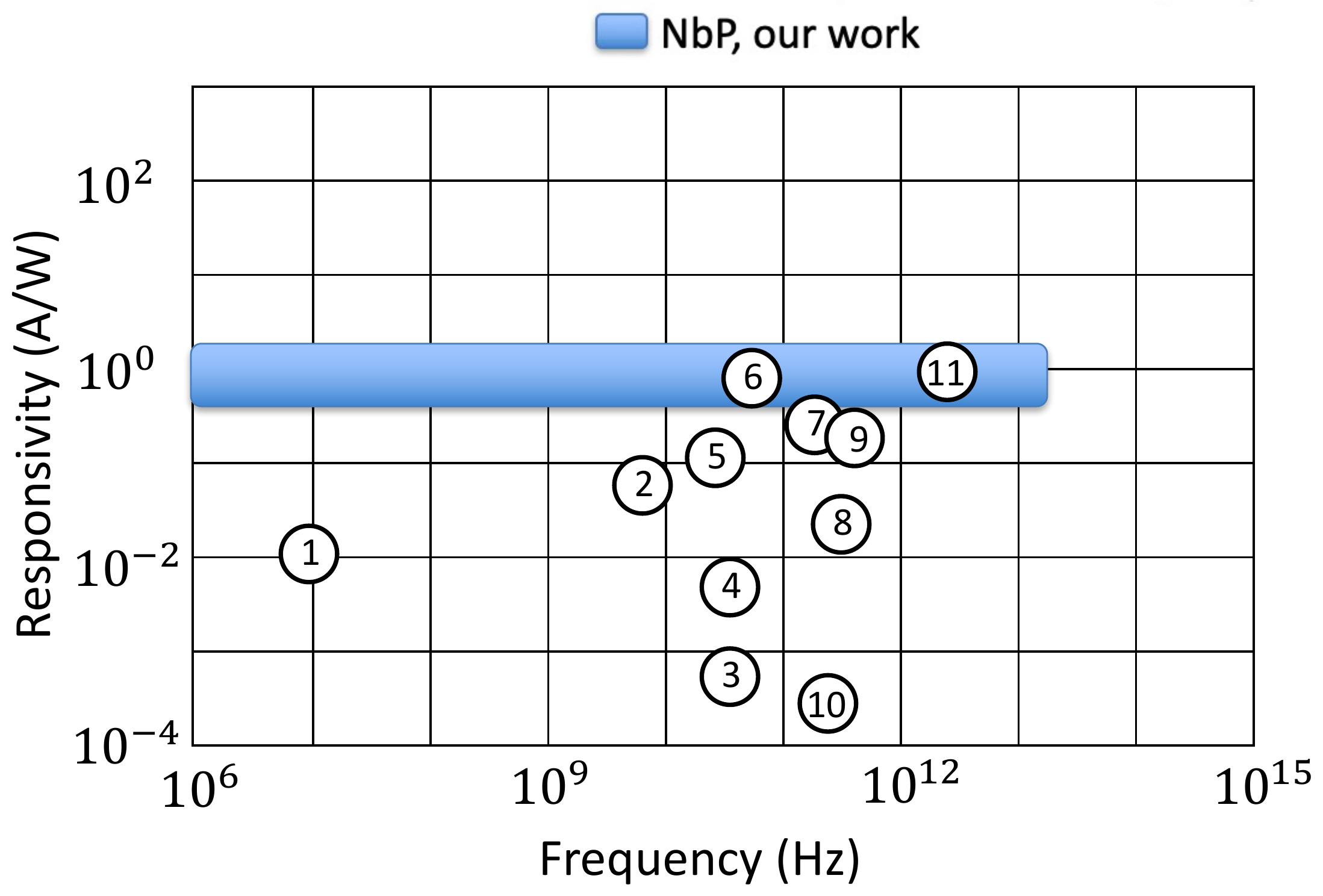}
\caption{
Comparison of current responsivity with previous study in the frequency range $10^6$ to $10^{15}$ Hz. The reference data points are adopted and updated from terahertz detector review article\cite{rogalski2019two}, label 1-11 corresponds to a series of work on two dimensional materials and topological semimetals \cite{yao2014high,pospischil2013cmos,xia2009ultrafast,schall201450,gan2013chip,cakmakyapan2018gold,guo2020anisotropic,auton2017terahertz,viti2016plasma,viti2015black,bandurin2018resonant}.
}
\label{fig2}
\end{figure}

Interestingly, we find that large contributions to Berry curvature dipole in NbP come from regions near two extended loops in momentum space, rather than the vicinity of Weyl points as in TaAs. These loops are located on a $\mathcal{M}_x$-invariant plane and correspond to a nodal line of band crossings that would occur due to band inversion if spin-orbit coupling were turned off. The introduction of spin-orbit coupling leads to strong momentum-dependent band hybridization and therefore creates anisotropic energy dispersion and large Berry curvature dipole near the Fermi level. As a result, a large current responsivity $R$ is found for a range of Fermi energy within $\pm 20$meV. Our calculation shows that the responsivity of NbP-based long-wavelength photodetector at zero external bias can reach about $0.3$A/W for a realistic device width $W=1\mu$m. Reducing the width to $W=0.1\mu$m leads to $R=3$A/W. We also calculate the current responsivity for other type-I Weyl semimetal TaAs, TaP and NbAs, type-II Weyl semimetal MoTe$_2$ and WTe$_2$ (NbAs shows a similar magnitude as NbP, see supplementary material for details).

In Fig. 3, we compare the performance of our nonlinear Hall long wavelength photodetector with other detectors and rectifiers. To this end, we plot their current responsivity versus operating frequency \cite{rogalski2019two,yao2014high,pospischil2013cmos,xia2009ultrafast,schall201450,gan2013chip,cakmakyapan2018gold,guo2020anisotropic,auton2017terahertz,viti2016plasma,viti2015black,bandurin2018resonant}. In the long wavelength region below 37 THz, the current responsivity is typically limited to less than $1$ A/W and varies dramatically with operating frequency. As the Drude half-width of NbP at room temperature determined from optical conductivity is around 30 meV \cite{neubauer2018optical}, we expect our nonlinear Hall photodetector can operate up to 15 THz with a large current responsivity reaching  $3$ A/W.

Another important figure of merit for photodetection is the noise-equivalent power (\textit{NEP}), defined as the noise power density over the responsivity. At room temperature, thermal noise is the dominant noise source compared to shot noise from discrete incident photon. We compute the \textit{NEP} from the root mean square
of the noise current as $\textit{NEP}=\frac{\sqrt{4 k_B T/r_{0}}}{R}$, where $T$ is the operating temperature and $r_0$ is the sample resistance. With sample width $W=0.1\mu$m and room temperature resistance $r_0=300 (\Omega)$ \cite{shekhar2015extremely}, the \textit{NEP} of NbP device at peak responsivity is estimated to 2.5 pW/Hz$^{0.5}$. For comparison,  the \textit{NEP} of typical Schottky diode based terahertz detector is around 100 pW/Hz$^{0.5}$ at sub THz ($0.1\sim1$ THz) range, and quickly increases at higher frequency. 

In addition to the Berry curvature dipole, other mechanisms including skew scattering and side jump also contribute to the nonlinear Hall effect \cite{Isobe,Spivak,Xie,Xie2,Niu,Niu2,Spivak2,inti2,rudner2019self}. For crystals with certain point groups such as $C_{3v}$,  Berry curvature dipole is forbidden by symmetry. In this case, theory predicts that in diffusive systems, second-order nonlinear transport arises from skew scattering and side jump induced by the inherent chirality of Bloch electron wavefunction \cite{Isobe}.  On the other hand, in ballistic transport regime, nonlinear Hall conductance is entirely determined by the integral of Berry curvature over half of the Fermi surface \cite{papaj2019magnus}, which is also an intrinsic property of Bloch electron wavefunction. Regardless of the microscopic origins, the inherent second-order nonlinearity of quantum materials can be utilized for long-wavelength photodetection by rectifying oscillating electric field into direct current.

On the experimental side,
second-order response was observed in recent electrical transport \cite{he2020quantum} and THz photocurrent experiments \cite{olbrich2014room,plank2016photon} on topological insulators. This effect arises from the Dirac surface states with hexagonal warping \cite{fu2009hexagonal}, which results in Berry curvature and skew scattering. The dependence of photocurrent on incident frequency between 0.5 and 4 THz  is consistent with Drude behavior expected from intraband processes.
Moreover, a very recent experiment achieved room temperature THz detection and imaging using the intrinsic second-order nonlinear response of topological semimetal PdTe$_2$ \cite{guo2020anisotropic}, which comes from the surface. The photocurrent as  a function of incident power displays excellent linearity. Remarkably, the current responsivity  without bias reaches as large as $0.2$A/W.
These exciting results on many fronts encourage future development of terahertz/infrared technology based on Hall rectifiers.

\begin{acknowledgements}
The work was supported by the U.S. Army Research Laboratory and the U.S. Army Research Office through the Institute for Soldier Nanotechnologies, under Collaborative Agreement Number W911NF-18-2-0048. YZ and LF were supported in part by Simons Investigator Award from the Simons Foundation.
\end{acknowledgements}

\bibliography{ref}

\end{document}